\documentclass[a4paper,11pt]{article}
\pdfoutput=1 

\usepackage{jinstpub}
\newcommand{\D}{\,\mathrm{d}}
\newcommand{\E}{\mathrm{e}}
\title{\boldmath Kinetics of relativistic electrons undergoing small recoils in periodic structures and matter}

\author[a,1]{E. Bulyak,\note{Corresponding author.}}
\author[a,b]{N. Shul'ga}

\affiliation[a]{NSC KIPT, 1, Academicheskaya Street, Kharkov UA-61108, Ukraine}
\affiliation[b]{V.N. Karazin National University,
Svobody Square, 4, Kharkov UA-61022, Ukraine}

\emailAdd{bulyak@kipt.kharkov.ua}

\abstract{We developed a general method for evaluating the energy spectrum evolution of relativistic charged particles that have undergone small quantum losses, such as the ionization losses when the electrons pass through matter and the radiation losses in the periodic fields. These processes are characterized by a small magnitude of the recoil quantum as compared with the particle's initial energy. The `detector' function for arbitrary recoil spectrum is derived in addition to the straggling function. These functions are determined by the average number of scattering events undergone by the particle over the length of the detector (ionization losses) or the periodic field and the specific spectrum of the recoil. Moments for both the straggling function and the detector function are derived. The minimum average number of recoils, above which both functions depend only upon the mean energy  of recoil quantum, is estimated. The average of this number is about 10 recoil events.}


\arxivnumber{1711.09571} 

\proceeding{XII$^{\text{th}}$ International Symposium <<Radiation from Relativistic Electrons in Periodic Structures>>\\
September 18--22, 2017\\
Hamburg, Germany}

\notoc
\begin{document}
\maketitle

\flushbottom

\section{Introduction}
Relativistic charged particles passing through the matter or (periodic) field, suffer a degradation of their energy due to various interaction processes. In some of the processes, namely the ionization losses in matter and the radiation losses in periodic structures --- channeling radiation in crystals, Compton inverse radiation in laser pulses or radiation in undulators --- portion of the energy lost in each interaction is small as compared to the initial energy of the particle. These physical processes are widely employed in the detectors of the charged particles, e.g., silicon detectors \cite{bichsel88} (ionization losses) and the so-called `laser wire monitors' \cite{laserwire05} (Compton scattering).

Study of the classical problem --- the spectrum evolution of the charged particles undergone the ionization losses --- began many decades ago, see \cite{bichsel88,shulga96} and the references therein. Degradation of the energy of the electron beams due to radiation in the periodic structures has been under study since the end of twentieth century, see \cite{khokonov04e,kolchuzhkin03,petrillo13,smolyakov14,bulyak17nimb}.
Such a process may be attributed to the quantum radiation effects below the radiation dominance, see \cite{piazza10}.

Formally, the problem setup was the following. A relativistic charged particle traversing a uniform matter or a periodic field, loses a small fraction of the initial energy $\gamma_0$ randomly in the form of small but definite independent portions with the maximum recoil energy $\omega_\mathrm{max}$: $\omega_\mathrm{max}\ll \gamma_0$, $x\omega_\mathrm{max}\ll \gamma_0$ with $x$ being the average number of scattering events over a fixed path length. Here and below we use the reduced energy units: $\gamma$ is the energy of the charged particle, $\epsilon$ is the energy in the straggling and the detector spectra, $\omega$ is the energy of the spectrum of the recoil quantum, all are dimensionless, normalized to the energy unit, e.g., to the charged particle rest energy.

The initial spectrum $f_0 (\gamma)$ of a beam of charged particles having traversed the media of thickness $\Delta $ is transformed into $f_\Delta (\gamma)$. These two spectra are connected with the so-called `straggling function' $S_\Delta(\epsilon)$ via cross-correlation operation:
\begin{align} \label{eq:initial}
f_\Delta(\epsilon) &= \left(f_0 \star S_\Delta\right)(\epsilon) \;,\\
\intertext{where $\star $ is the cross correlation operator, }
\left(g\star h\right)(y) &\equiv\int_{-\infty}^{\infty} g(x+y) h(x)\D x\; .\nonumber
\end{align}

The goal of the experiments is to derive the third function from the two known ones. The problem is that the `detector', e.g. a bulk of matter traversed by the beam or a target irradiated by the beam emitted x-rays, registers the signal $G_\Delta(\epsilon)$ which in general does not coincide with the straggling function.

In this paper we consider the relation of the detector function to the straggling function.

\section{Kinetics of the particles and the straggling function}
Specific features of the `small' losses are: (i) small magnitude that allows to consider them independent of particles' energy; (ii) compact support of the recoil spectra: with  the minimum loss magnitude of $\omega_\text{min}\ge 0$, and the maximum loss of $\omega_\text{max}$. Thus, the spectrum of losses (the differential cross-section as in \cite{bichsel88,bichsel06}) has a compact support, $\omega_\text{min}\le \omega\le \omega_\text{max}$, and can be normalized
\begin{equation}\label{eq:normal}
 \int_{-\infty}^{\infty} W(z,\omega)\D\omega < + \infty\qquad \to \qquad W(z,\omega ) = \frac{W(z,\omega)\int_{-\infty}^{\infty} W(z,\omega)\D\omega}{\int_{-\infty}^{\infty} W(z,\omega)\D\omega}=\psi(z)w(\omega)\; ,
\end{equation}
where $z$ is the coordinate along the axis of the system.

The normalization \eqref{eq:normal} is equivalent to the presentation of the energy losses as statistically independent quantum recoils with the probability of $\psi (z)$ and the basic spectrum of $w(\omega)$, see \cite{bichsel88,bulyak17nimb}.

\subsection{Kinetics of the charged particles}
As we have shown in our recent paper \cite{bulyak17nimb}, under the condition of small losses,  a solution to the balance equation \cite{landau44,vavilov57},
\begin{equation*}
\frac{\partial f_x(\gamma)}{\partial x}  = \int_{-\infty}^\infty w(\omega) \left[  f_x(\gamma + \omega)- f_x(\gamma )\right]\D\omega \, ,
\end{equation*}
(also known as the transport equation, \cite{bichsel88}) --- the characteristic function of the spectrum evolution --- is
\begin{equation} \label{eq:charactfun}
\hat{f}_x = \hat{f}_0 \E^{x(\check{w}-1)}\; ,
\end{equation}
where $f_0(\gamma ) = f_{x=0}(\gamma)$ is the initial spectrum, $x=\int_{0}^{z}\psi(\zeta )\D\zeta $ is the mean number of collisions from the entry to the system up to the coordinate $z$, $\hat{g}\equiv \mathcal{F}\{g\}$ is the Fourier transform of the function $g$ and $\check{g}\equiv \mathcal{F}^{-1}\{g\}$ is the inverse Fourier transform.

As was shown in \cite{bulyak17nimb}, this solution may be presented as the Poisson-weighted sum of the self-states $F_n$ (the spectrum taken in exact $n$ recoils), see \cite{bichsel88,khokonov04e}:
\begin{subequations}\label{eq:crosscor}
\begin{align}\label{eq:poiweight}
f_x(\gamma) &= \sum_{n=0}^\infty \frac{\mathrm{e}^{-x} x^n}{\Gamma(n+1)} F_n(\gamma)\; ;
\\
F_n(\gamma)&= (w\star F_{n-1})(\gamma)\;,\qquad F_0(\gamma)=f_0(\gamma )\; . \label{eq:croself}
\end{align}
\end{subequations}

%
%

The straggling function, normalized to unity, presents the loss spectrum: only the particles that have undergone at least one recoil contribute to it.
The characteristic function for the straggling function and its Poisson-weighted expansion are
\begin{align} \label{eq:fourstra}
\hat{S}_x  &=  \hat{w}\E^{x(\hat{w}-1)}\; , &
S_x(\epsilon )&= \sum_{n=0}^\infty \frac{\mathrm{e}^{-x} x^n}{\Gamma(n+1)} F'_n(\epsilon)\; ,\\
F'_n&=F'_{n-1}\ast w\; , & F'_0&=w\; . \nonumber
\end{align}
where $\ast $ stands for the convolution operation, $(g\ast h)(y) = \int_{-\infty}^{\infty} g(x) h(y-x)\D x$, see \cite{bichsel88}.
Comparing \eqref{eq:fourstra} with \eqref{eq:crosscor}, one may see that $F'_n(\epsilon ) = F_{n+1}(\gamma_0-\epsilon)$.

\subsection{The detector function}
Implying a linear detector that collects the losses over the interval $[0,x]$, we get the detector function (normalized to unity):
\begin{equation}\label{eq:detect}
\hat{G}_x = \frac{\hat{w}}{x(\hat{w}-1)}\left[\E^{x(\hat{w}-1)}-1\right]\; ;  \qquad
G_x(\epsilon) = \frac{1}{x} \sum_{n=0}^{\infty}\left[ 1-\frac{\Gamma(1+n,x)}{\Gamma(1+n)}\right]F'_n(\epsilon)\; ,
\end{equation}
where $\Gamma(1+n,x)$ is the (upper) incomplete gamma function.
The self-states $F'_n(\epsilon)$  are the same as in the straggling function \eqref{eq:fourstra}.

\begin{figure}
\centering
  \includegraphics[width=0.49\textwidth]{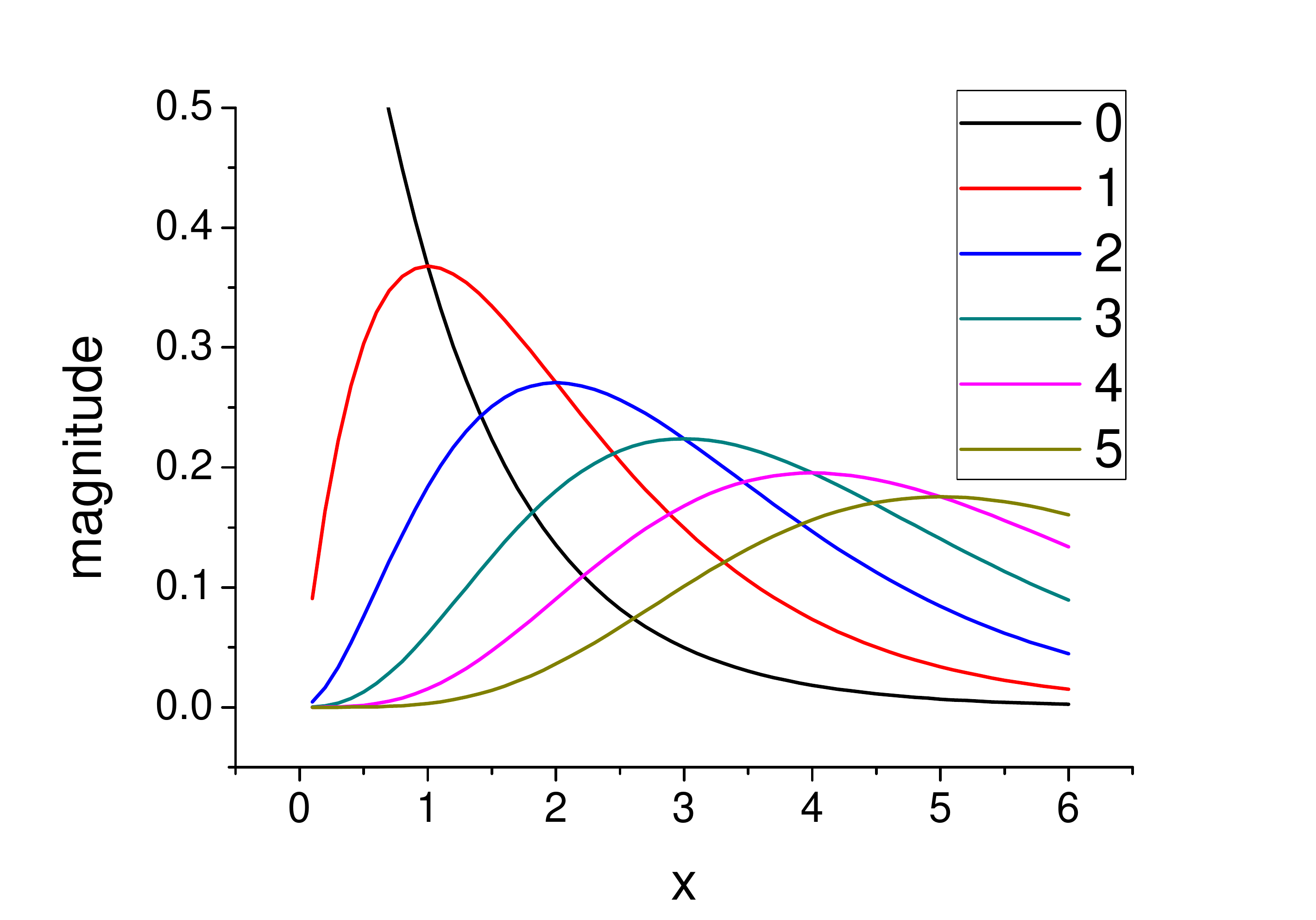}
  \includegraphics[width=0.49\textwidth]{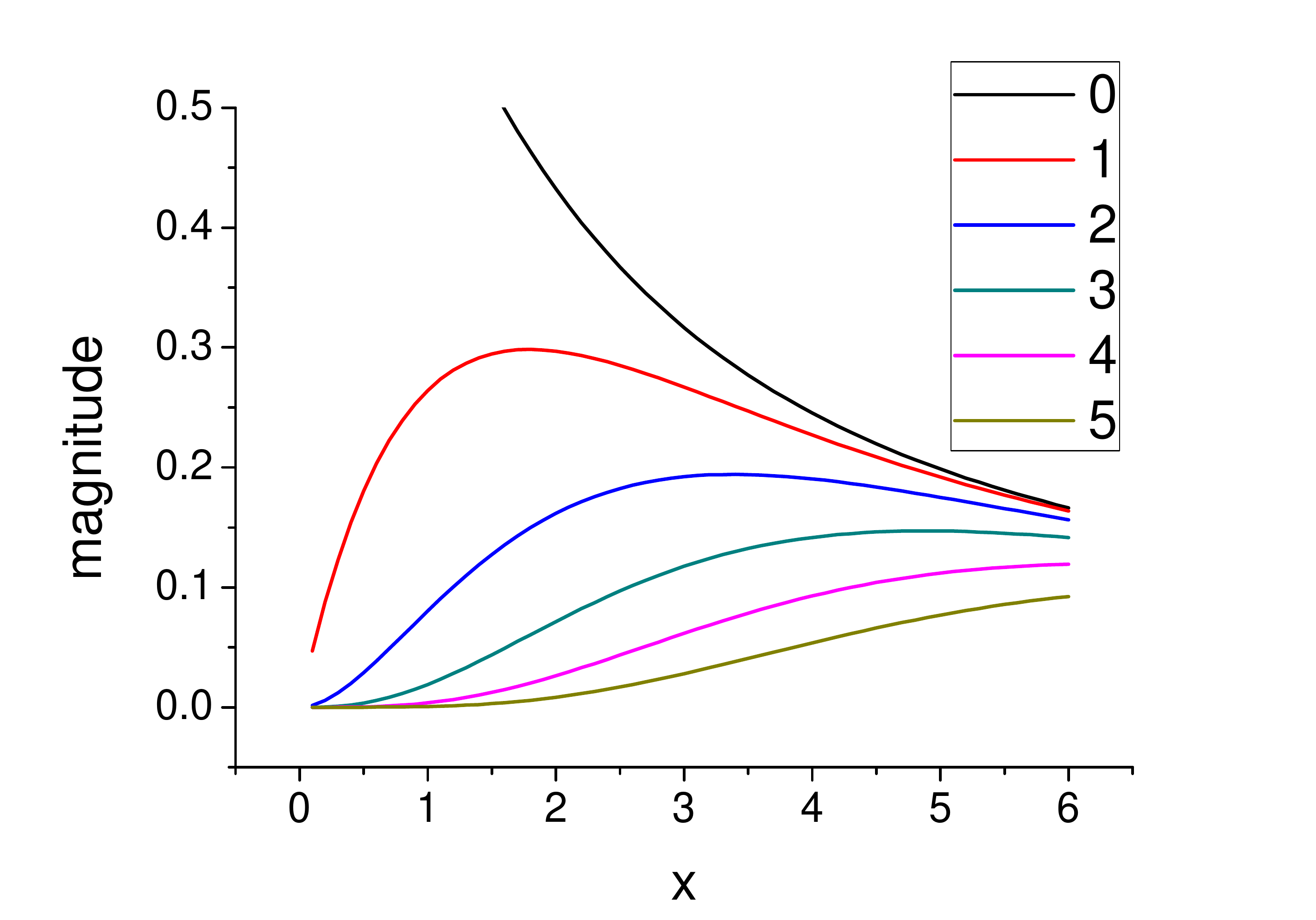}
\caption{\label{fig:weight} Poisson weights (left) and integral weights (right) vs. average number of recoils  for $n=0,\cdots,5$.}
\end{figure}

Comparing the straggling function \eqref{eq:fourstra} with the detector function \eqref{eq:detect}, we should emphasize the different weights with which the self states contribute to them: the Poisson weights for the straggling spectra and the `integral Poisson' for the detector function, as is depicted in figure\,\ref{fig:weight}.

As can be seen from figure\,\ref{fig:weight}, the Poisson weight reaches its maximum at $n=x$ for $n>0$, while the straggling weight reaches maximum much later.

\subsection{Moments}
Following the procedure described in \cite{bulyak17nimb}, we can derive the moments of the detector function \eqref{eq:detect} and the straggling function \eqref{eq:fourstra}.
The first three  moments --- the mean energy $\overline{\epsilon}$, the variance $\mathrm{Var}[\epsilon]\equiv \overline{\left(\epsilon - \overline{\epsilon}\right)^2}$ and the skewness $\mathrm{Sk}[\epsilon]\equiv \overline{\left(\epsilon - \overline{\epsilon}\right)^3}$ --- for both the detector function (left column) and the straggling  spectrum (right column, see \cite{bulyak17nimb}) --- are:
\begin{subequations}\label{eq:rawmom}
\begin{align}\label{eq:1raw}
  \overline{\epsilon} &=\left(1+\frac{x}{2}\right) \overline{\omega}\; , &  \overline{\epsilon} &= (1+x)\, \overline{\omega}\; ;\\
  \mathrm{Var}[\epsilon] &=\left(1+\frac{x}{2}\right) \overline{\omega^2} +\left(\frac{x^2}{12}-1\right) \overline{\omega}^2\; , & \mathrm{Var}[\epsilon] &=(1+x)\, \overline{\omega^2}-\overline{\omega}^2\; ;\label{eq:2raw} \\
  \mathrm{Sk}[\epsilon] &= \left(1+\frac{x}{2}\right) \overline{\omega^3} +\left(\frac{x^2}{4}-3\right) \overline{\omega^2}\, \overline{\omega}+2\,\overline{\omega}^3\; , & \mathrm{Sk}[\epsilon] &= (1+x)\, \overline{\omega^3}-3\, \overline{\omega^2}\, \overline{\omega} +2\, \overline{\omega}^3  \; . \label{eq:3raw}
\end{align}
\end{subequations}
Here $\overline{\omega}$ is the mean energy and $\overline{\omega^2}$ is the dispersion (the second raw moment) of the basic recoil spectrum $w(\omega)$.

As can be seen from \eqref{eq:rawmom}, the moments of the straggling function and the detector function --- being equal to each other and to that of the recoil spectrum at $x=0$ --- behave in different ways with increase of $x$ (the detector `thickness'). The mean energy, $\overline{\epsilon}$, is smaller for the detector function than the straggled, almost twice at $x\gg 1$; the variance (width) of the detector function is smaller for $x < x_\ast $ and larger above $x_\ast $. The quantity,
\begin{equation} \label{eq:xast}
x_\ast = 6 \overline{\omega ^2}/\overline{\omega} ^2\, ,
\end{equation}
may be regarded as a limit above which the detector function does not convey information on the particular shape of the recoil spectrum. For physically possible recoils this parameter gets the value: $6\le x_\ast \le 12$.

\section{Ionization losses and radiation in periodic structures}
Both the straggling spectrum and the detector function depend on the energy through the same set of the self-states; the difference is the weights, which are determined by the average number of the recoils.

\subsection{Ionization losses}
\begin{figure}
\centering
  \includegraphics[width=0.49\textwidth]{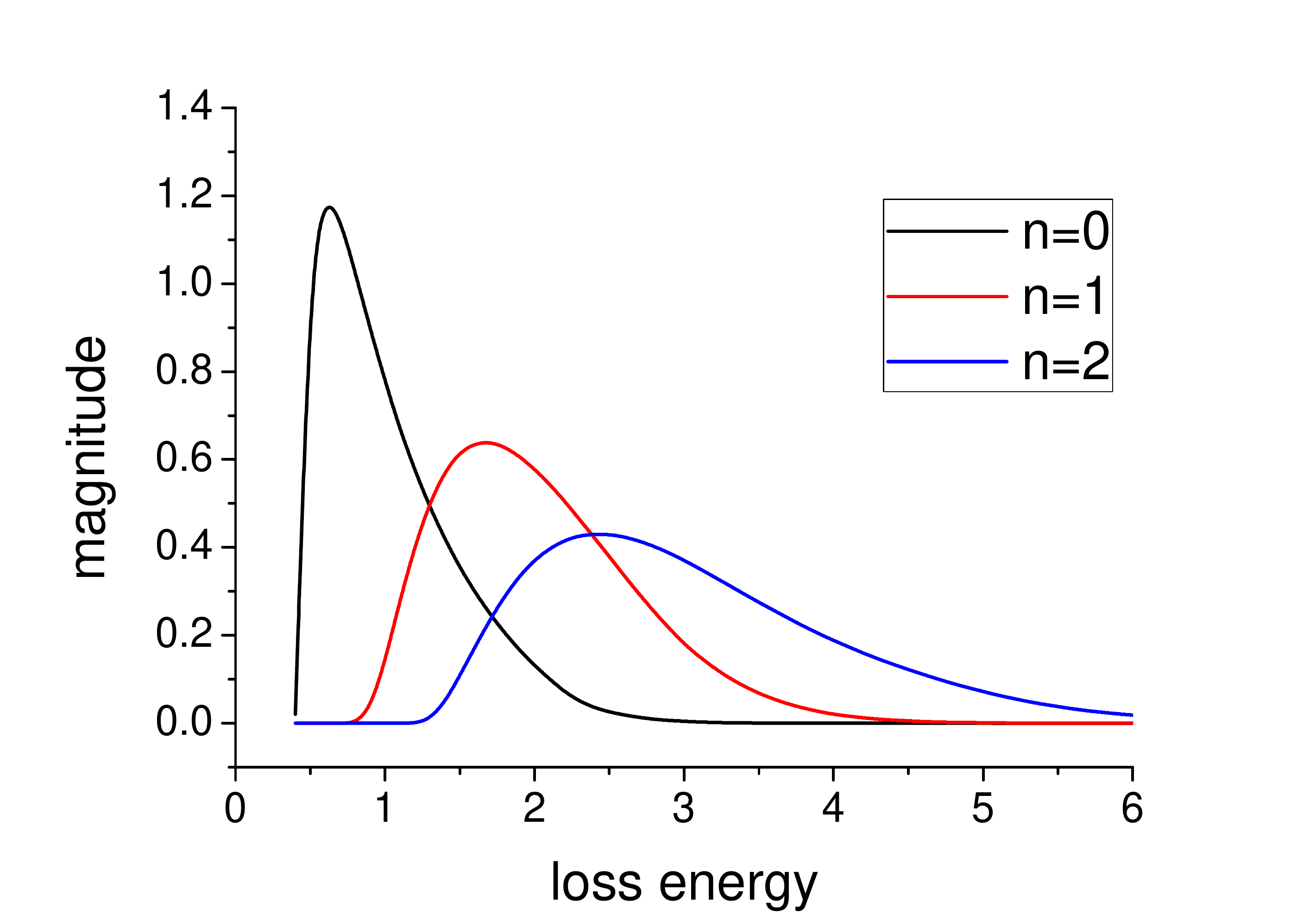}
  \includegraphics[width=0.49\textwidth]{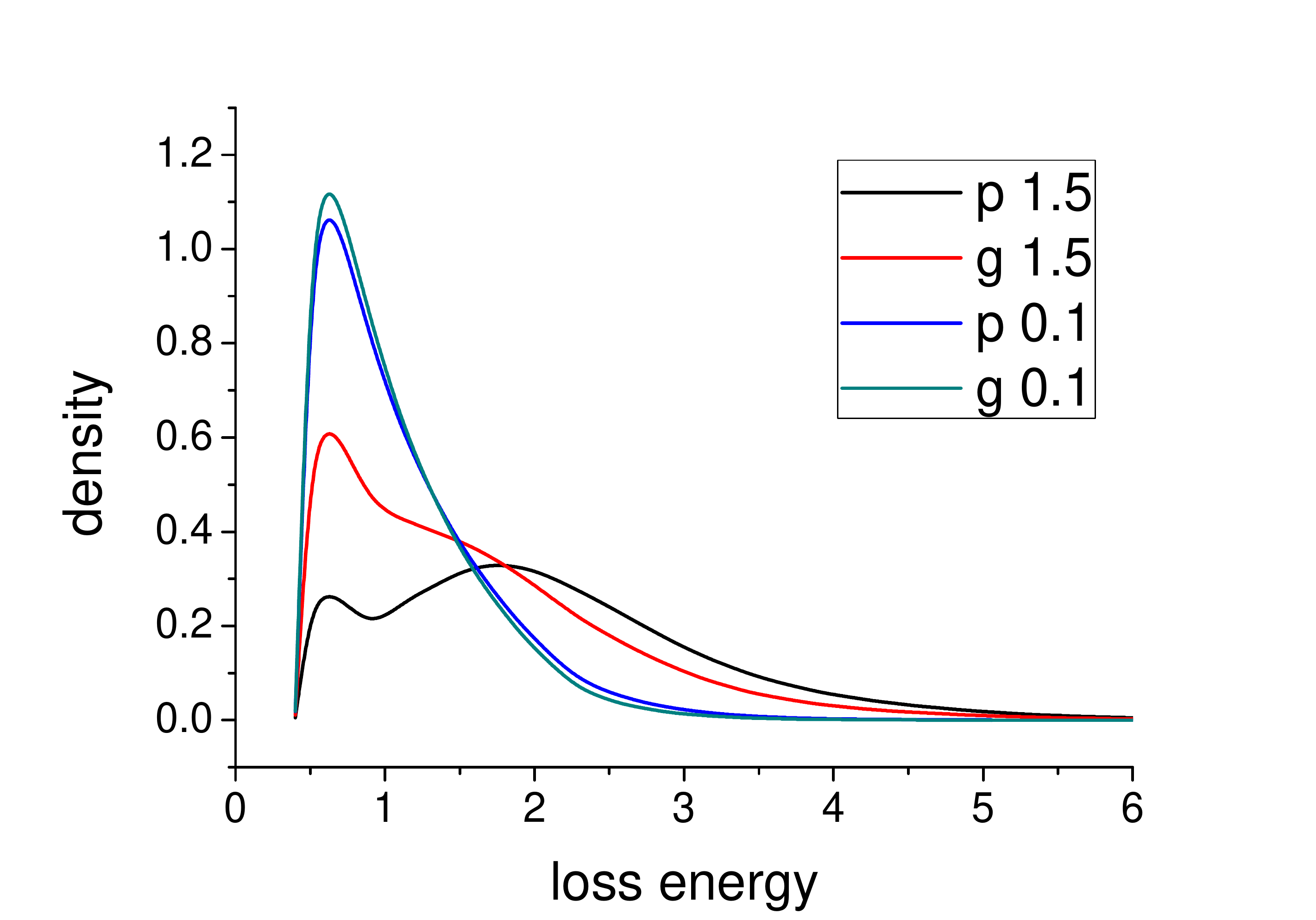}
\caption{\label{fig:incomp} Left: Self states, right: Poisson-weighted (straggling) distribution (p) and `detector' distribution (g) for $x=0.1,1.5$.}
\end{figure}

We illustrate the developed method with a qualitative study of the ionization losses. Lets consider a simplified recoil spectrum extended from the minimum to the maximum energy, $\omega_\mathrm{min}\le \omega\le \omega_\mathrm{max}$, with the declining dependence on the energy loss, $w(\omega )\propto \omega^{-2}$, see \cite{landau44,vavilov57}. We then study a smoothed spectrum depicted in figure\,\ref{fig:incomp}, the self-state $n=0$.

Dependence of contributions of the first four self-states on the cumulative spectrum of the average number of scattering is presented in figure\,\ref{fig:incomp}. The case of $\omega_\mathrm{min} = 0.4$, $ \omega_\mathrm{max} = 4$ is considered.

A general characteristic of the self states, which stem from the convolution (cross correlation) procedure, is the following: $n$ self-state has compact support $[n \omega_\mathrm{min}, n\omega_\mathrm{max}]$, i.e., each subsequent state is wider and smoother than previous, see also \cite{bulyak17nimb} for the dipole radiation recoils.
For the particular dependence $\propto \omega^{-2}$ of the basic function $w(\omega)$ (zero self-state) $n$th self state behaves as $\propto \omega^{-(n+2)}$.

The low-energy part of the detector function does not change its shape with the increase in the thickness of the radiator: it builds up with $x$ to saturation and the tail of the function extends $\sim \omega_\mathrm{max}(1+x)$.  From the physical point of view, the tail length remains finite since there is a finite maximum number of recoils, $n_\mathrm{max}<\infty $, due to the finite number of the electrons with which the charged particle can interact over a finite pass length (or a finite number of the driving field periods).

\subsection{Radiation from the periodic structures}
Specifics of the periodic structures are the following: (i) the periodic structure may be the quasi-periodic, with the field strength varying slowly such as in laser pulses or tapered undulators and (ii) the `detector' registers only a small fraction of the emitted radiation spectrum (in most cases along the charged particle average trajectory).

The first feature, quasi-periodicity, implies a nonlinear dependence of the average number of recoils $x$ on the physical length of the trajectory $z$: the function $\psi (z) $ in eq. \eqref{eq:normal} therefore is not constant.

A consequence of the second feature is that the straggling function resembles the \emph{particle} spectrum in the primarily employed case  of the pin-hole collimation of the detector placed collinearly with the averaged particle trajectory, see \cite{bulyak14a}.

In this case the detector function is integral of the kinetic spectrum \eqref{eq:crosscor} with $f_0 = \delta(\omega - \omega_\mathrm{max})$, where $\omega_\mathrm{max}$ is the maximum energy of the photons emitted by the electron with the initial energy $\gamma_0$.

Respectively, the moments of the detector function, which in this case yields the spectrum   of radiation, read:
\begin{equation}\label{eq:dipmean}
\overline{\epsilon} = \omega_\mathrm{max}\left( 1 - \frac{x \overline{\omega}}{2\gamma_0}\right)\; , \quad
\mathrm{Var}[\epsilon] = \frac{x}{2}\overline{\omega^2}+\frac{x^2}{12}\overline{\omega}^2\; .
\end{equation}

For the dipole radiation, emitting from the undulator with small deflection parameter or small-power laser pulse, the detected variance becomes equal to the beam spectrum variance at the average number of scattering events $x_\ast = 8.4 $. This quantity defines the limit above which the specific shape of the recoil spectrum becomes not distinctive in the detected spectrum.

A similar to \eqref{eq:dipmean} expression for the mean energy of the multiphoton  channeling radiation was obtained in \cite{bondarenco15} with different technique.

\section{Conclusion}
A general method for the evaluation of the straggling function for arbitrary loss spectrum was proposed and considered.
It was shown, that the different known methods for computation of the straggling function (see the report \cite{bichsel88}) --- the transport equation, the convolution method, the method of moments --- have the same basis: the transport equation (also referred to as the balance equation). This equation describes the evolution of the spectrum of the charged particle that loses its  energy in small fractions.

The particle is considered the source of the signal registered by the detector. Thus the detector collects the signals produced by the evolving particle spectrum from the entire volume of the detector itself (ionization losses) or the periodical field (radiation losses). It was found that the evolution of the spectrum as well as the detected signal was determined by the spectral shape of the individual recoil quantum.
The derived characteristic functions for the evolution of the charged particle spectrum and the detected signal are determined as the Fourier transform of the recoil spectrum and the average number of the recoils.

Both the straggling spectrum of the particles and the detected (integral) spectrum can be presented as a weighted sum of the self-states. The self-states are the same for the straggling and the detected spectra, but the weights are different: the Poisson ones for former and the integral Poisson for the latter.
The moments of both the straggling and the detected signal are evaluated for the arbitrary recoil spectrum. It should be emphasized that the description of the distribution density function with the moments has a physical meaning for relatively thick radiator, $x> x_\ast$, while a particular shape of the recoil spectrum does not matter.

\acknowledgments
This work is partially supported by the Ministry of Education and Science of Ukraine, project No 1-13-15. Participation of the authors in RREPS 2017 Symposium and presentation of the report was partially supported by DESY laboratory.

\providecommand{\noopsort}[1]{}\providecommand{\singleletter}[1]{#1}

\providecommand{\href}[2]{#2}\begingroup\raggedright\endgroup
\end{document}